\documentclass[]{jingdong}
\definecolor{jsonstr}{rgb}{0.5,0.5,0.5}
\usepackage{authblk} 

\usepackage{marvosym}
\usepackage{listings}
\usepackage{xcolor}
\usepackage{booktabs}
\usepackage{colortbl}  
\usepackage{xcolor}    
\lstdefinelanguage{json}{
    basicstyle=\ttfamily\tiny,
    showstringspaces=false,
    breaklines=true,                 
    breakatwhitespace=false,
    frame=single,                    
    rulecolor=\color{gray!40},
    backgroundcolor=\color{gray!5},  
    literate=
     *{:}{{{\color{black}{:}}}}{1}
      {,}{{{\color{black}{,}}}}{1}
      {\{}{{{\color{black}{\{}}}}{1}
      {\}}{{{\color{black}{\}}}}}{1}
      {[}{{{\color{black}{[}}}}{1}
      {]}{{{\color{black}{]}}}}{1},
    morestring=[b]",
    stringstyle=\color{jsonstr},
}

\usepackage[utf8]{inputenc}
\usepackage[T1]{fontenc}

\usepackage{amsfonts}
\usepackage{amsmath}
\usepackage{amssymb}
\usepackage{algorithm}
\usepackage{algpseudocode}
\usepackage{booktabs}
\usepackage{xcolor}
\usepackage{enumitem}
\usepackage{inconsolata}
\usepackage{listings}
\usepackage{nicefrac}
\usepackage{siunitx}
\usepackage{url}
\usepackage{xspace}
\usepackage{wrapfig}
\usepackage{graphicx}

\crefformat{section}{\S#2#1#3}
\Crefformat{section}{\S#2#1#3}
\crefmultiformat{section}{\S#2#1#3}{ and \S#2#1#3}{, \S#2#1#3}{ and \S#2#1#3}
\Crefmultiformat{section}{\S#2#1#3}{ and \S#2#1#3}{, \S#2#1#3}{ and \S#2#1#3}
\crefrangeformat{section}{\S#3#1#4 to \S#5#2#6}
\Crefrangeformat{section}{\S#3#1#4 to \S#5#2#6}

\definecolor{templatekeyword}{HTML}{E1251B}
\definecolor{templatestring}{HTML}{0B7A75}
\definecolor{templatecomment}{HTML}{6B7280}
\definecolor{templatecodebg}{HTML}{F7F7F8}

\lstdefinestyle{templatecode}{
  basicstyle=\ttfamily\small,
  columns=fullflexible,
  backgroundcolor=\color{templatecodebg},
  frame=single,
  rulecolor=\color{black!15},
  numberstyle=\tiny\color{gray},
  keywordstyle=\color{templatekeyword},
  commentstyle=\color{templatecomment},
  stringstyle=\color{templatestring},
  showstringspaces=false,
  tabsize=2,
  breaklines=true,
  breakatwhitespace=true,
  captionpos=b,
  xleftmargin=3.4pt,
  xrightmargin=3.4pt
}

\title{JoyAI-Talker: Full-Duplex Speech Interactive Large Model Built for Empathetic Voice Agents}

\vspace{-40pt}

\affiliation[]{JD.com}

\vspace{-40pt}

\abstract{
We present \textbf{JoyAI-Talker}, a full-duplex speech dialogue system that delivers robust foundation model capabilities while empowering \textbf{empathetic interaction and voice agent intelligence}. 
JoyAI-Talker adopts a modular Thinker-Talker architecture that decouples cognitive planning, conversational state coordination, and speech generation, and further implements a unified speech-text joint training pipeline to mitigate the common "cognitive degradation" bottleneck, thereby largely preserving the model’s core textual reasoning, STEM, and logical capabilities while extending them to speech-based interaction.
For expressive speech synthesis, the Talker module employs a text-controllable generation paradigm that enables natural-language instructions to flexibly control vocal attributes and localized paralinguistic events, such as laughter and sighs, supporting more expressive and fine-grained speech responses.
To enhance conversational empathy, we introduce the \textbf{Persona-Adaptive Empathetic Response (PAER)} framework. PAER employs a hierarchical cognitive pipeline to extract non-verbal speaker cues, such as gender, age, and emotional state, from raw input audio, incorporate them into the Thinker's CoT reasoning, and generate context-adaptive responses that align semantically appropriate text with fine-grained control over utterance-level expressiveness and localized paralinguistic events, including sighs, speaking rate, and volume. We further integrate \textbf{Joy-Duplex}, a state-driven, plug-and-play full-duplex framework that functions as an efficient gating engine for real-time turn control. 
Extensive evaluations show that JoyAI-Talker achieves \textbf{highly competitive performance on foundational text-to-text (T2T) and speech-to-text (S2T) benchmarks}. In full-duplex evaluation, the system reaches a high response rate of $0.88$ under user interruptions while maintaining an extremely low false-trigger rate under background speech, demonstrating its readiness for fluid and natural speech dialogue.

\vspace{-30pt}
}

\begin{document}

\maketitle

\begin{figure}[] 
\vspace{-27pt}
  \centering
  \begin{subfigure}[b]{0.32\textwidth}
    \centering
    \includegraphics[width=\textwidth]{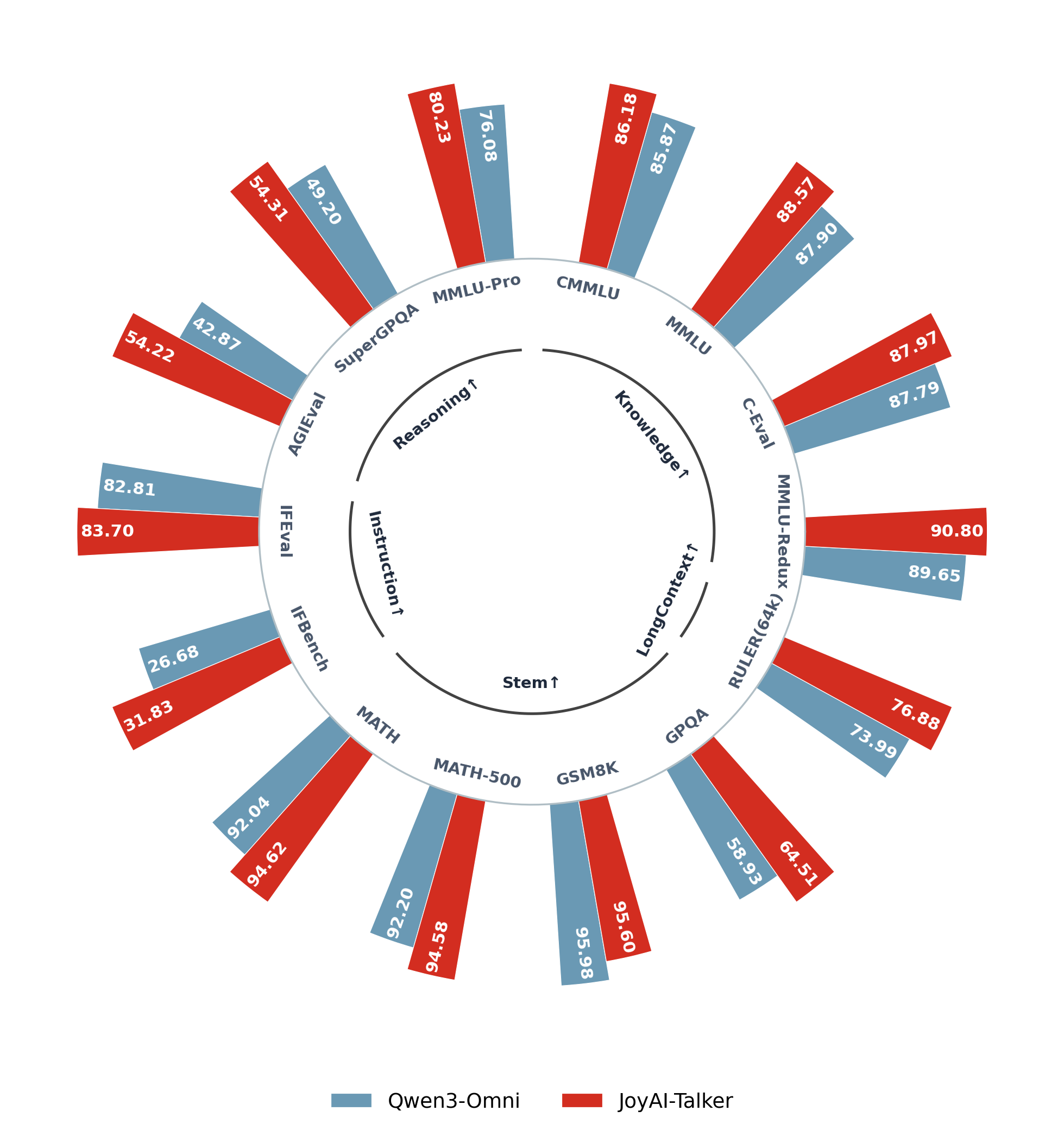} 
    \vspace{-15pt}
    \caption{T2T} 
    \label{fig:sub_a} 
  \end{subfigure}
  \hfill 
  \begin{subfigure}[b]{0.32\textwidth}
    \centering
    \includegraphics[width=\textwidth]{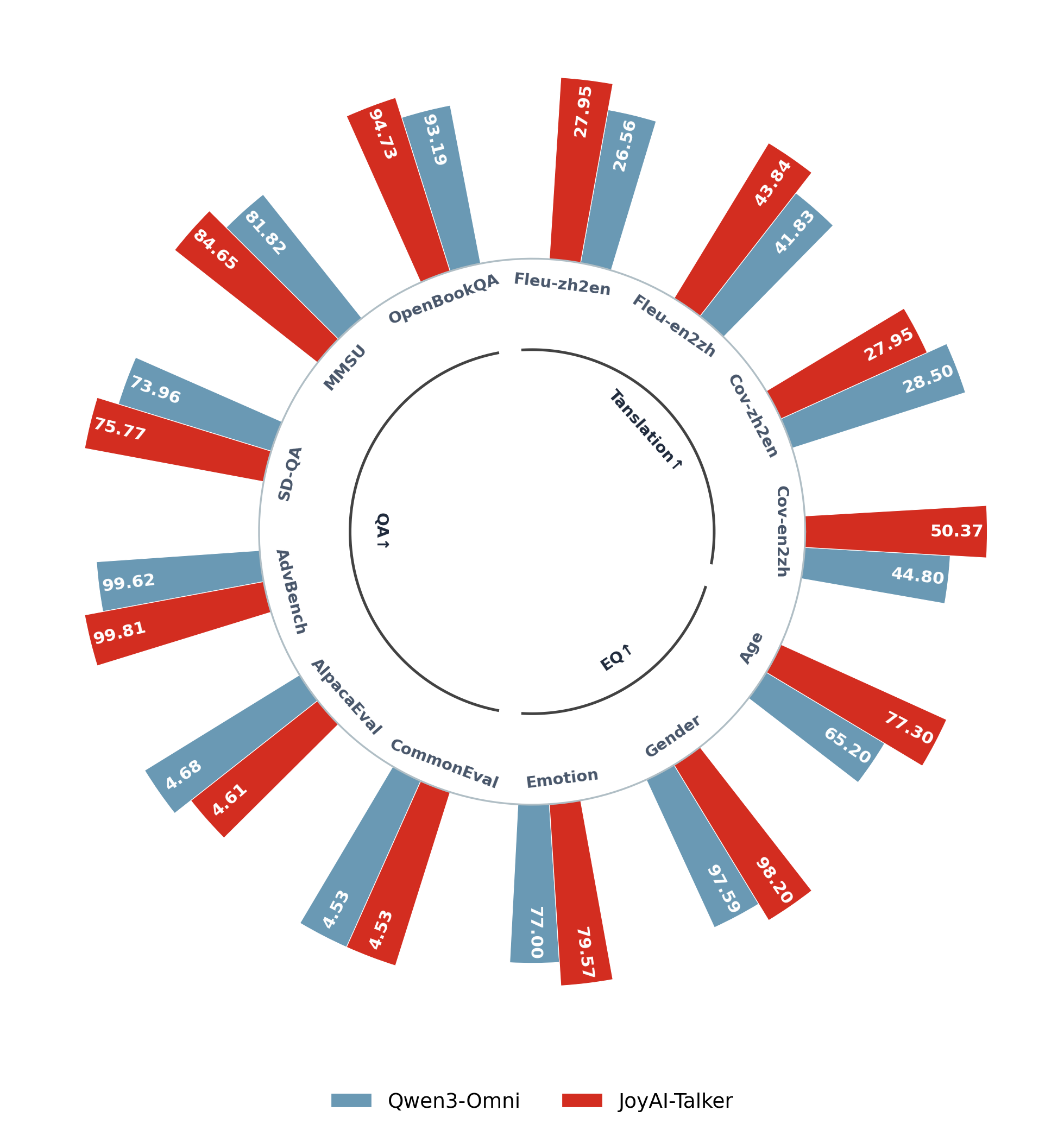}
    \vspace{-15pt}
    \caption{S2T - QA, Tanslation, EQ} 
    \label{fig:sub_b}
  \end{subfigure}
  \hfill 
  \begin{subfigure}[b]{0.32\textwidth}
    \centering
    \includegraphics[width=\textwidth]{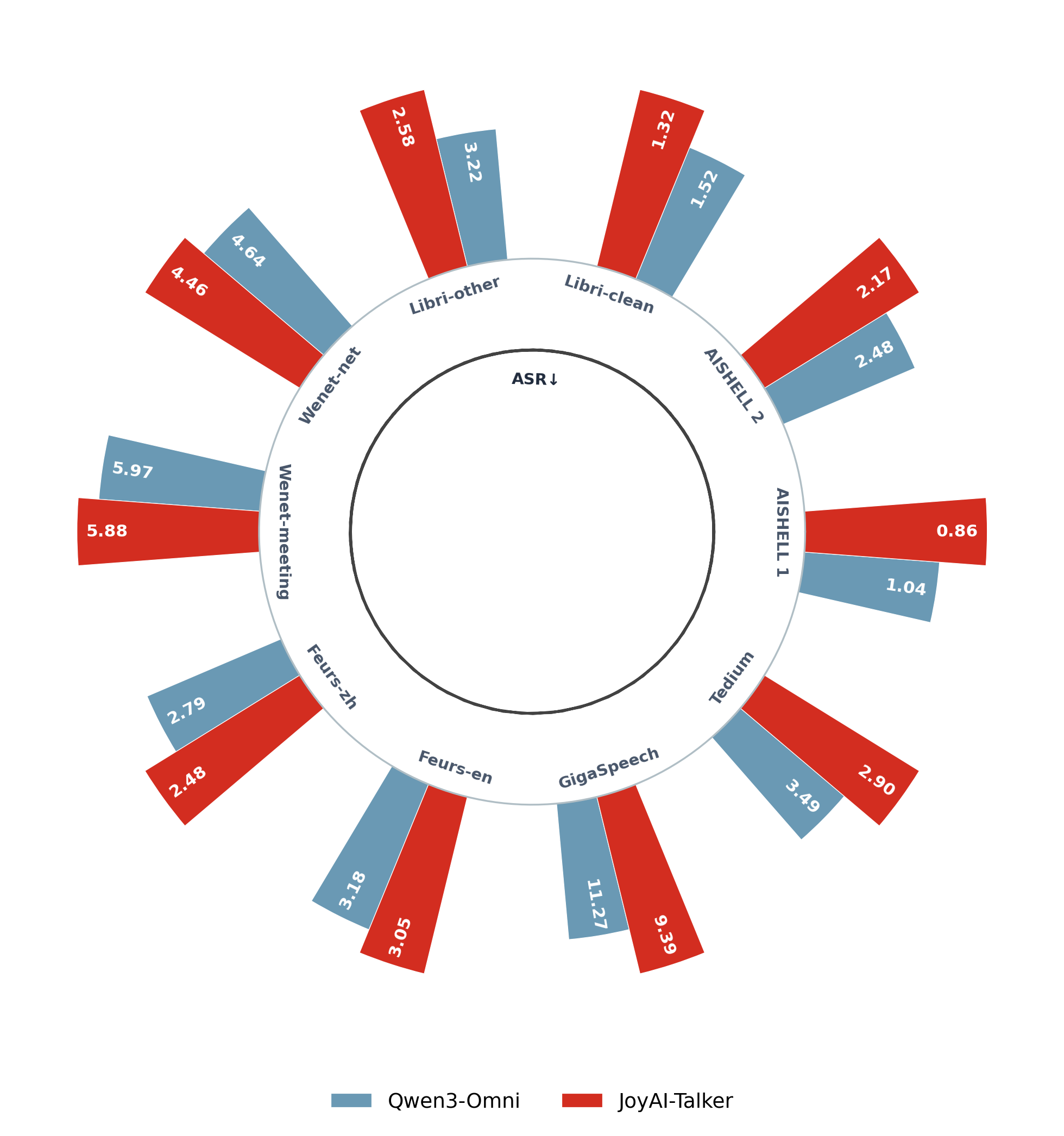}
    \vspace{-15pt}
    \caption{S2T - ASR} 
    \label{fig:sub_c}
  \end{subfigure}
  \vspace{-8pt}
  \caption{Overall performance of JoyAI-Talker across T2T and S2T benchmark.}
  \label{fig:main_figure} 
\end{figure}

\section{Introduction}
\label{sec:introduction}

The rapid advancement of Large Language Models (LLMs) has profoundly accelerated the development of intelligent conversational agents. However, since speech remains the most natural, expressive, and efficient modality for human communication, the research paradigm is increasingly shifting from text-centric models toward Large Speech Language Models (LSLMs). Historically, speech dialogue systems relied on cascaded architectures—comprising Automatic Speech Recognition (ASR), an LLM, and Text-to-Speech (TTS) modules. While functional, these pipelines suffer from compounded latency and the irreversible loss of rich paralinguistic cues (e.g., emotion, intonation, and hesitation) during the text-bottleneck conversion. Recently, the industry has witnessed a breakthrough in end-to-end multi-modal architectures. Advanced proprietary models, such as GPT-4o and Gemini 3.1 Live, have pushed the boundaries of real-time, highly expressive voice interactions. Concurrently, open-weight pioneers like Moshi~\cite{defossez2024moshi}, Freeze-Omni~\cite{freezeomni2024}, and Qwen3-Omni~\cite{qwen3omni2025} have explored various dual-stream or natively unified structures to achieve lower latency and richer acoustics. Despite these unprecedented strides, achieving truly human-like, immersive, and intelligent conversational agents still faces three fundamental challenges:

\begin{itemize}
    \item \textbf{Modality-Induced Cognitive Degradation:} Integrating the continuous, high-dimensional speech modality into the discrete linguistic space of LLMs often triggers ``catastrophic forgetting.'' During joint multi-modal training, models typically sacrifice their foundational logical reasoning, mathematical proficiency, and complex problem-solving capabilities, forcing a suboptimal trade-off between textual intelligence and acoustic alignment.
    
    \item \textbf{Lack of Dynamic and Contextual Empathy:} Current voice assistants tend to generate flat, generalized responses. While some models can simulate basic emotions, they struggle to dynamically perceive fine-grained speaker attributes (e.g., age, gender, and underlying emotional states) from the input audio. Intertwining logical reasoning with rich paralinguistic behaviors (e.g., laughter, sighs, and hesitations) remains difficult, making it challenging to deliver the customized, empathetic interactions characteristic of genuine human communication.
    
    \item \textbf{Fragile Full-Duplex Interaction:} Real-world conversations are highly unstructured, filled with overlaps, backchannels, and ambient noise. Traditional full-duplex systems~\cite{defossez2024moshi} or energy-based Voice Activity Detection (VAD) methods suffer from high false activation rates. As observed in recent benchmarks~\cite{lin2026fullduplexv1.5}, speech dialogue agents tend to adopt extreme policies: they are either over-reactive (triggering heavily on background noise and interruptions) or over-conservative (failing to yield during genuine user speech), thereby disrupting the natural conversational flow.
\end{itemize}

To address these critical bottlenecks, we introduce \textbf{JoyAI-Talker}, a full-duplex speech dialogue system that delivers expressive, empathetic, and stable real-time full-duplex interactions. JoyAI-Talker adopts an innovative, decoupled \textbf{Duplex-Thinker--Talker} architecture. The language backbone is built upon JoyAI-LLM Flash~\cite{cai2026joyaillmflashadvancingmidscale}, a 48.9B-parameter sparse Mixture-of-Experts (MoE) model featuring Multi-Head Latent Attention (MLA) and auxiliary-loss-free routing, inspired by recent large-scale MoE architectures such as DeepSeek-V3~\cite{deepseekai2025deepseekv3technicalreport} and Kimi-K2~\cite{kimiteam2026kimik2openagentic}. The \textit{Thinker} focuses on acoustic understanding, empathetic reasoning, and response formulation, while the \textit{Talker}—which follows the speech generation architecture of JoyVoice~\cite{joyvoice}—acts as a controllable, low-latency speech generator capable of interpreting natural-language instructions and explicit paralinguistic tokens.

The key features and capabilities of JoyAI-Talker are summarized as follows:

\begin{itemize}
    \item \textbf{Foundational Capability for Conversational Voice Agents:} 
    Instead of functioning as a narrow interaction model, JoyAI-Talker is developed as a highly versatile foundation model that delivers competitive performance across a diverse range of cognitive, reasoning, and linguistic benchmarks. By maintaining a unified semantic and acoustic modeling interface, the system simultaneously provides a robust, extensible base to support conversational voice agents, enabling multi-turn speech reasoning and tool-use.
    
    \item \textbf{Unified Joint Training with Alleviated Cognitive Degradation:} 
    Rather than delaying speech-text integration until the post-training phase, we adopt a unified speech-text joint training paradigm starting from the early mid-training phase. By carrying this joint modeling through Mid-training, Context Extension, SFT, and DPO, JoyAI-Talker alleviates the common "cognitive degradation" bottleneck to a certain extent. The model preserves its fundamental reasoning, STEM, and logical capabilities in the textual domain (e.g., reaching 94.62\% on MATH), demonstrating a favorable balance between text-based intelligence and speech interaction.
    
    \item \textbf{Persona-Adaptive Empathetic Response (PAER) via CoT:} 
    We introduce a progressive cognitive empathy framework structured as a hierarchical pipeline: audio understanding $\rightarrow$ Chain-of-Thought (CoT) reasoning $\rightarrow$ empathetic dialogue generation. Guided by this pipeline, the Thinker first extracts speaker attributes—such as gender, age, and emotional state—from the raw input audio. It then integrates these perceived attributes into its CoT reasoning to formulate context-adaptive responses. These responses combine semantically appropriate text with precise, token-level controls over utterance-level expressions (e.g., speaking rate and volume) and localized paralinguistic events, allowing the Talker to deliver expressive, personalized speech while maintaining a consistent assistant identity.
    
    \item \textbf{Decoupled, Text-Controllable Expressive Speech Generation:} 
    The Talker module employs a decoupled, instruction-controllable speech generation paradigm to translate semantic outputs and speaking instructions into speaker-consistent, expressive speech. By separating response planning from acoustic execution, the Talker dynamically controls prosody, emotional expression, and localized paralinguistic behaviors (e.g., \texttt{[Laughter]} and \texttt{[Sigh]}) via natural-language instructions and explicit vocabulary tokens, while maintaining a stable assistant identity.
    
    \item \textbf{Joy-Duplex: A Non-Intrusive, Highly Stable Full-Duplex Framework:} 
    Departing from computationally heavy end-to-end parallel dual-stream architectures, we propose a modular, state-driven, plug-and-play full-duplex framework. By integrating fine-grained interaction state tokens directly into the streaming decoding path, Joy-Duplex serves as an efficient gating engine for real-time turn control. On the rigorous Full-Duplex-Bench v1.5~\cite{lin2026fullduplexv1.5}, Joy-Duplex delivers a well-balanced interaction policy, achieving a high responsiveness of 0.88 under user interruptions while keeping false triggers extremely low under background speech, representing a robust and stable interaction policy for fluid speech dialogue.
\end{itemize}

\begin{figure}[htbp]
\centering 
\includegraphics[width=0.85\textwidth]{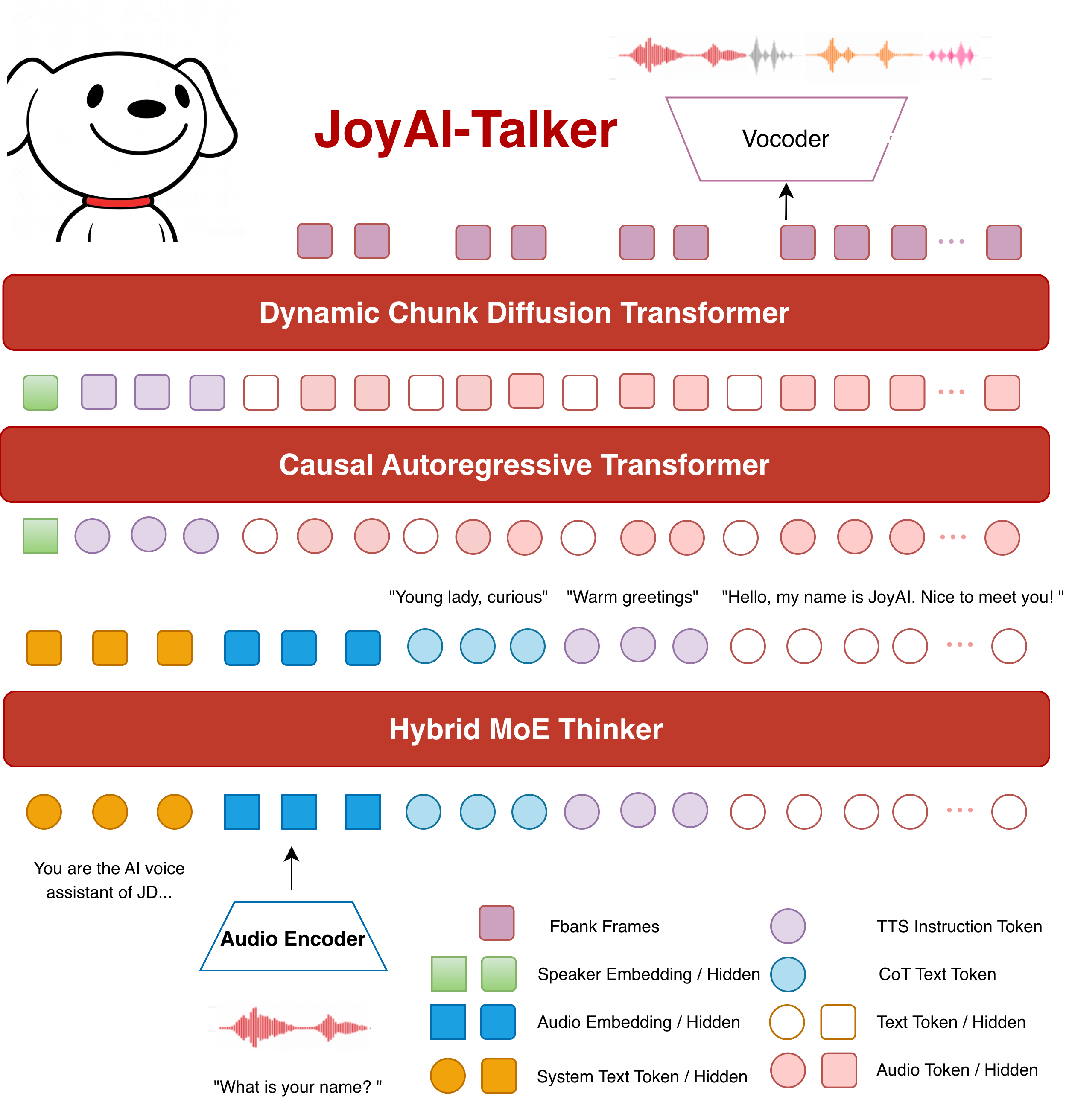}
\caption{An overview of Thinker-Talker pipeline in JoyAI-Talker. It consists of several key components: 
1) Audio Encoder: This module processes the input speech query and extracts continuous audio embeddings. 
2) Hybrid MoE Thinker: Ingesting multimodal inputs, this cognitive core performs reasoning to predict the target text tokens and TTS instruction tokens. 
3) Causal Autoregressive Transformer: This module takes the speaker embeddings, predicted TTS instruction tokens, and target text tokens as input, and predicts the discrete audio tokens. 
4) Dynamic Chunk Diffusion Transformer: Using the hidden representations from the Causal Autoregressive Transformer as input, this component predicts the mel-spectrogram output. 
5) Vocoder: This module is responsible for reconstructing the predicted mel-spectrogram back to continuous waveform audio.} 
\label{fig:example} 
\end{figure}
\section{Duplex-Thinker-Talker Architecture}
JoyAI-Talker is designed as a decoupled, state-driven, full-duplex speech dialogue system structured around three core pillars: the \textbf{Duplex Interaction}, the \textbf{Thinker}, and the \textbf{Talker}. Unlike computationally heavy, end-to-end parallel dual-stream architectures, this modular paradigm separates conversational state coordination, cognitive semantic reasoning, and expressive acoustic synthesis. By establishing a clear separation of concerns, the system achieves a balanced trade-off between semantic intelligence, acoustic expressiveness, and engineering deployability.

\paragraph{Architectural Advantages:}
This decoupled modular design offers two major advantages:
\begin{itemize}
    \item \textbf{Modularity and Independent Optimization:} Each component can be trained, fine-tuned, or upgraded independently without retraining the entire end-to-end pipeline. This plug-and-play capability greatly simplifies engineering maintenance and reduces development costs.
    
    \item \textbf{Preservation of Cognitive Capabilities:} In tightly coupled speech dialogue systems, real-time interactive responsiveness often conflicts with deep semantic intelligence. To enable low-latency full-duplex interaction, models must constantly process fine-grained temporal steps, whether through real-time frame prediction over short temporal chunks~\cite{thinkingmachines2026interactionmodels} or continuous dual-stream audio token ingestion as in Moshi-like architectures~\cite{defossez2024moshi}. Integrating these high-frequency, low-level streaming classification tasks directly into the primary language backbone fragments its contextual attention and dilutes its parameter capacity, leading to severe cognitive degradation and a drop in overall reasoning performance. By offloading these rapid, short-chunk state-tracking and interactive routing tasks to the outer Duplex Interaction Layer, our design shields the Thinker's semantic core, allowing it to focus exclusively on coherent, long-context reasoning and empathetic alignment.
\end{itemize}

\paragraph{Limitations and Mitigations:}
Nonetheless, the decoupled paradigm introduces certain engineering and architectural trade-offs:
\begin{itemize}
    \item \textbf{Slightly Higher Latency Overhead:} Sequential routing across modular boundaries (even when optimized via chunk-based streaming) naturally introduces a minor latency penalty, resulting in a slightly higher turn-taking latency than unified, single-pass parallel models.
    
    \item \textbf{Information Bottleneck at the Interface:} The symbolic interface between the Thinker and the Talker prevents the direct propagation of continuous acoustic nuances (such as the user's exact pitch fluctuations or micro-hesitations from the input). We actively mitigate this limitation by generating structured, localized paralinguistic tokens (e.g., \texttt{[Sigh]}, \texttt{[Laughter]}) and emotion-aware speaking instructions. These explicit cues allow the Talker to reconstruct rich emotional prosody in a highly controllable and precise manner.
    
    \item \textbf{Infrastructure Complexity:} Managing asynchronous full-duplex interactions—such as real-time barge-in, voice activation detection (VAD), and stream cancellation—requires a relatively complex, state-driven serving architecture.
\end{itemize}

\subsection{Thinker: Empathetic Voice-Agent Speech-Text Foundation Model}

JoyAI-Talker adopts a standard encoder–projector–decoder architecture, integrating the JoyAI-LLM Flash language model with a speech encoder through lightweight MLP projectors. An overview of the model architecture is presented in Figure~\ref{fig:example}.

For speech understanding, the speech encoder follows an attention-based encoder–decoder architecture that has been pre-trained on large-scale automatic speech recognition (ASR) corpora. Input audio is first converted into log Mel filter-bank features and then downsampled by a factor of 8× using stacked Conv2D blocks before being processed by the Transformer encoder. This design reduces the acoustic token rate to 12.5 Hz, significantly improving computational efficiency while preserving the semantic information required for downstream language modeling.

The language backbone is built upon JoyAI-LLM Flash~\cite{cai2026joyaillmflashadvancingmidscale}, a sparse Mixture-of-Experts (MoE) large language model comprising 48.9B total parameters, with approximately 3.28B parameters activated for each input token. Following recent large-scale MoE architectures such as DeepSeek-V3~\cite{deepseekai2025deepseekv3technicalreport} and Kimi-K2~\cite{kimiteam2026kimik2openagentic}, the backbone incorporates Multi-Head Latent Attention (MLA) together with RMSNorm, Rotary Position Embeddings (RoPE), and SwiGLU feed-forward networks.

The backbone consists of 40 Transformer layers, including one dense Transformer layer followed by 39 sparse MoE layers. Each MoE layer contains a fine-grained pool of 256 experts and adopts an auxiliary-loss-free load-balancing routing strategy. For each input token, the router dynamically activates the Top-8 routed experts together with a dedicated shared expert, achieving a favorable balance between model capacity and computational efficiency.

\subsubsection{Training Recipe}

Initialized from the JoyAI-LLM Flash pretraining phase, JoyAI-Talker undergoes a unified, multi-stage speech-text joint training paradigm across all training phases. Our motivation for initiating joint multi-modal training at the mid-training phase—rather than delaying it until the SFT phase—is twofold. First, the early foundational pretraining stages establish the model's fundamental cognitive baseline and world knowledge over massive textual corpora. Second, integrating the speech modality at this earlier stage promotes stable cross-modal alignment and prevents the catastrophic forgetting of textual knowledge, thereby improving multi-modal understanding without inducing textual degradation. 

Consequently, we implement a unified speech-text mixed training scheme across all subsequent phases, sequentially optimizing the model through four distinct stages:

\begin{itemize}
    \item \textbf{Speech-Text Joint Mid-training:} This phase aims to bridge the representation space of the speech and textual modalities while preserving the foundational cognitive capabilities of the pre-trained language backbone. We introduce a highly diversified joint corpus. For the textual modality, this comprises web crawls, reasoning-intensive code repositories, high-fidelity PDF documents, and synthetic data to balance general knowledge with logical reasoning. For the speech modality, we incorporate a wide variety of tasks including automatic speech recognition (ASR), speech translation, speech question-answering (QA), audio captioning, real-world multi-turn speech dialogues, and speech text continuation (synthesized by feeding ASR transcriptions into an LLM to generate coherent textual responses paired with the original audio). This joint pretraining is split into two speech-text joint steps: 
    \begin{itemize}
        \item \textit{Step 1 (Audio Adapter Warmup):} Both the LLM backbone and the audio encoder are frozen, and we optimize only the multi-modal audio adapter to map high-dimensional speech features into the linguistic embedding space of the LLM.
        \item \textit{Step 2 (Full-Parameter Joint Training):} We unfreeze all model parameters—including the LLM backbone, the speech encoder, and the audio adapter—for comprehensive end-to-end joint optimization.
    \end{itemize}
    Throughout this phase, speech data is formatted in a chat-style Q\&A structure, while textual data is ingested as raw, continuous text to preserve cognitive capacity.
    
    \item \textbf{Speech-Text Joint Context Extension:} Following mid-training, this phase is designed to comprehensively bolster reasoning capabilities over extensive textual documents and enhance interactive proficiency during extended, multi-turn speech dialogues. All model parameters remain fully trainable. To target general long-context scaling, the textual mixture is rebalanced by upsampling long-context mathematical and coding datasets. For the speech modality, we upsample real-world multi-turn long speech dialogues and dynamically synthesize multi-turn ASR and translation tasks by concatenating single-turn samples. To emphasize long-range dependencies while preserving short-context proficiency, the joint data mixture blends long-form text (academic papers, financial reports, etc.) and long-duration audio (podcasts, lectures, etc.) with a balanced proportion of short-sequence data.
    
    \item \textbf{Speech-Text Joint Supervised Fine-Tuning (SFT):} Serving as the cornerstone of our post-training alignment pipeline, this phase is strategically designed to expand the model's multi-modal knowledge boundaries while cultivating highly colloquial, concise, and direct speech communication. The textual SFT dataset covers code, math, STEM, and general instruction-following, where we upsample knowledge-dense and interaction-centric categories to facilitate communicative voice dialogues. The speech SFT dataset covers ASR, translation, QA, speech text continuation, audio captioning, and empathetic speech dialogues. To facilitate cross-modal alignment and elevate speech response quality, we upsample synthesized speech-to-text (S2T) data constructed by converting conversational textual SFT prompts into natural speech audio via our in-house text-to-speech model.
    
    \item \textbf{Speech-Text Joint Direct Preference Optimization (DPO):} To further refine response quality, improve instruction-following, and suppress residual hallucinations, we implement a dedicated DPO phase. The model is optimized directly on pairwise preferred and dis-preferred responses self-generated by the SFT policy. For each seed prompt (spanning general QA, reasoning, and speech inputs), the SFT policy generates multiple candidate responses, from which preferred and dis-preferred pairs are selected using a multi-faceted model judging framework. Candidates are routed to three complementary, specialized judges:
    \begin{itemize}
        \item \textit{General QA Judge:} An LLM judge evaluates mathematics, code, STEM, and general knowledge responses along correctness, logical soundness, clarity, and completeness, using reference answers as quality anchors.
        \item \textit{Instruction-Following (IF) Judge:} For prompts with explicit structural constraints (exact item counts, mandatory keywords, length limits, JSON output), we prompt an LLM judge with a constraint-compliance rubric. This rubric allocates most points to hard-constraint satisfaction (with fixed penalties per violation type) and the rest to content quality and output cleanliness, thereby raising the candidate scoring margin on constraint-violating cases.
        \item \textit{Rule-Based IF Judge:} For machine-verifiable constraints, we score candidates using deterministic, rule-based validations, ensuring full reproducibility and zero inference cost.
    \end{itemize}
    For speech inputs, each prompt carries an audio field while the candidate responses remain textual, allowing us to leverage the same QA and IF judges on the transcribed text. Crucially, rather than pooling speech-conditioned and text-conditioned candidates cross-modally—which would create severe distribution mismatches and an excessively large optimization gap due to the systematic modality gap—we adopt a \textit{speech-only strategy}. This strategy constructs preference pairs strictly within the speech-conditioned pool. This local comparison provides a highly calibrated, realistic, and learnable gradient signal, enabling the model to steadily correct its speech-pathway behavior. The resulting textual and speech preference sets are then mixed for joint DPO training.
\end{itemize}

\subsubsection{Training Details}
Our audio-language model training system is built on a highly optimized extension of the Megatron-Core framework.   Alongside Data Parallelism (DP), Tensor Parallelism (TP), Sequence Parallelism (SP), and Pipeline Parallelism (PP), we employ 8-way Expert Parallelism (EP) to scale the sparse Mixture-of-Experts backbone.   MoE operations are accelerated through DeepEP for low-latency token dispatch and combination, grouped GEMM, and fused permutation and routing.   CUDA Graph capture further reduces CPU launch overhead and stabilizes step latency.

The training sequence length is configured at 8K tokens during the Mid-training Phase, and is subsequently scaled to 64K tokens during the Context Extension and SFT Phases.   For the 64K-token training stages, we adopt Context Parallelism (CP) with THD-format attention to distribute the computational load.   Under this long-context regime, we employ a best-fit sequence packing strategy utilizing block-diagonal attention masks to concatenate multiple independent samples or speech dialogue sessions into a single training sequence.   This is powered by an audio-aware packing algorithm that jointly considers token length and audio-padding waste, preventing cross-contamination between unrelated samples while approximately doubling end-to-end throughput over standard length-only packing.   We also distribute the audio encoder across CP ranks: each rank processes its local audio slice, and a differentiable all-gather reconstructs the embeddings while preserving gradient flow.   This reduces per-rank encoding cost in proportion to the CP degree and avoids an encoder bottleneck in long-context multimodal training.

Across all joint training phases, we implement a strict loss masking strategy to optimize representation learning for target generations.   For both textual corpora and multi-turn conversational datasets, the autoregressive loss is computed solely on the Assistant's responses.   The loss on all other components is completely masked out;   this includes all user-side inputs—encompassing both textual and speech queries—as well as all tool-side responses (such as tool execution results and API replies), on which no loss is computed.   This selective gradient optimization ensures that the model's representation learning capacity is focused exclusively on generating accurate, coherent, and concise assistant responses.

\subsubsection{Empathy}

Large Speech Language Models (LSLMs) have substantially advanced speech language understanding, driving applications ranging from intelligent assistants to empathetic companions \cite{DBLP:conf/naacl/WangZLSLZLAC25}. However, effective dialogue extends beyond merely interpreting \textit{what} is said, it necessitates recognizing speaker attributes (e.g., age and gender) and discerning \textit{how} the message is delivered \cite{DBLP:conf/iclr/ChengHYLF0J0Z0025, DBLP:conf/emnlp/YanLCNYMYC25}. Non-verbal acoustic cues-such as emotional prosody and paralinguistic behaviors (e.g., laughter, sigh, and cough)-are essential for capturing these nuances, facilitating natural, trustworthy, and emotionally intelligent communication. Despite their importance, existing approaches typically model these elements in isolation, failing to reflect the tightly coupled nature of understanding, reasoning, and response generation inherent in human conversation.

To address this gap, we introduce \textbf{Persona-Adaptive Empathetic Response (PAER)} to deliver highly customized, multifaceted empathic interactions that dynamically adapt to individual users. This framework is explicitly structured as a hierarchical cognitive pipeline: audio understanding $\rightarrow$ chain-of-thought (CoT) reasoning $\rightarrow$ empathetic dialogue generation. Our main contributions are twofold:

\begin{itemize}
    \item \textbf{A Progressive Cognitive Empathy Framework:} We design a progressive task formulation that mirrors human cognitive processes. Initially, we employ foundational perception tasks to accurately extract speaker attributes (e.g., gender, age, and emotional state) from the input audio. \textit{Thinker} module then explicitly integrates these attributes into its reasoning process via CoT. This facilitates the generation of context-adaptive responses that contain not
only contextually appropriate text but also precise controls over
utterance-level expression and localized paralinguistic events (e.g., sighing,
slow speaking rate, and low volume), without altering the assistant's speaker
identity. Subsequently, the Talker module, following the JoyVoice speech generation
framework~\cite{joyvoice} and refined through instruction-aware supervised
fine-tuning (SFT), translates the Thinker's textual and control outputs into
highly expressive and natural speech.
    
    \item \textbf{Comprehensive Evaluation and Superior Performance:} We rigorously evaluate our model across foundational audio understanding and downstream empathy generation tasks. First, we validate its precise perception of speaker attributes on public benchmarks, achieving outstanding performance on AIR-Bench \cite{DBLP:conf/acl/YangXLC0ZLLZZZ24} (for age and gender) and MER2025 \cite{DBLP:conf/mm/00040X0L000CZ0P25} (for emotion). These robust perceptual capabilities lay a solid foundation for individual-specific empathy. Building upon this, we comprehensively assess end-to-end empathic expressiveness using the open-source EchoMind dataset \cite{DBLP:journals/corr/abs-2510-22758}. Extensive experiments demonstrate that our approach consistently achieves superior performance in both textual and acoustic empathy metrics.
\end{itemize}

\subsection{Talker: Instruction-Controllable Expressive Speech Generation}
The Talker converts the textual response produced by the Thinker into
speaker-consistent, expressive, and streaming speech. It follows the
end-to-end Transformer--DiT architecture introduced in
JoyVoice~\cite{joyvoice}. An autoregressive Transformer predicts supervised
speech tokens, while its hidden representations directly condition a causal
Diffusion Transformer for acoustic generation. This architecture preserves the
stable linguistic modeling provided by discrete speech tokens while allowing
high-level textual and instructional information to propagate directly into
acoustic realization.

Beyond linguistic content, the Talker accepts natural-language speaking
instructions that specify how a response should be delivered. This design
separates response planning from speech realization: the Thinker determines
what to say and, when appropriate, produces a speaking instruction based on
its understanding of the user and dialogue context; the Talker realizes the
response in a fixed assistant voice with the intended emotion, prosody, vocal
effort, and localized paralinguistic behavior.

The Talker is obtained through instruction-aware foundation training followed
by speaker-specific supervised fine-tuning. Foundation training establishes
general speech generation, open-ended instruction following, paralinguistic
modeling, and streaming capabilities. Speaker-specific SFT then adapts these
capabilities to the target assistant voice. Consequently, speaker identity is
fixed in the deployed Talker, whereas utterance-level expressive attributes
remain dynamically controllable through local instructions.

\subsubsection{Training Data}

Following the large-scale speech data construction recipe of
JoyVoice~\cite{joyvoice}, the Talker is trained on a multilingual corpus
containing both read and in-the-wild speech. The corpus covers diverse
speakers, languages, recording conditions, speaking styles, and conversational
scenarios, providing broad support for robust pronunciation, natural prosody,
and generalizable speech generation.

To construct instruction-aware training data, we further collect expressive
speech from movies, television programs, and other conversational media.
Compared with conventional read speech, these sources contain richer variations
in emotion, speaking rate, vocal effort, conversational rhythm, phonation, and
paralinguistic behavior. Multimodal contextual cues from audio and video are
used to derive natural-language descriptions of speaker characteristics and
utterance-level acoustic realization.

The expressive corpus is combined with general speech data during foundation
training. General speech preserves linguistic coverage, pronunciation accuracy,
and default synthesis quality, while expressive speech expands the controllable
distribution of speaking styles and paralinguistic behaviors.

\subsubsection{Hierarchical Global and Local Instructions}

The instruction-controllable foundation model uses a hierarchical representation
consisting of global and local instructions. The global instruction describes
relatively stable speaker-level characteristics, such as role description,
gender, age, accent, and underlying timbre. The local instruction specifies
the acoustic realization of the current utterance, including its scene,
temporary timbre variation, primary emotion, speaking rate and rhythm,
breathiness, and spatial impression.

This separation distinguishes persistent speaker characteristics from
utterance-dependent expression. During speaker-specific SFT, the global
characteristics are absorbed into the fixed target-speaker model, while the
local instruction remains available as the dynamic control interface. Therefore,
the deployed Talker does not require a global instruction at inference time.

\subsubsection{Sequence Formulation}

Following the unified autoregressive formulation of
JoyVoice~\cite{joyvoice}, we extend the input sequence with hierarchical
instruction conditions:
\begin{equation}
\mathcal{I}
=
\left[
P;\,T;\,S
\right],
\end{equation}
where
\begin{equation}
\begin{split}
P &=
\left[
P_{\mathrm{sys}};\,
I^{g}
\right],\\
T &=
\left[
\mathrm{spk};\,
I^{l};\,
t_{1},t_{2},\ldots,t_{N}
\right],\\
S &=
\left[
s_{1},s_{2},\ldots,s_{M}
\right].
\end{split}
\end{equation}

Here, $P_{\mathrm{sys}}$ defines the speech generation task, $I^{g}$
provides persistent speaker-level conditioning, $\mathrm{spk}$ denotes the
speaker embedding, and $I^{l}$ specifies the local expressive characteristics
of the current utterance. The sequences $\{t_n\}_{n=1}^{N}$ and
$\{s_m\}_{m=1}^{M}$ denote the input text tokens and target speech tokens,
respectively.

Placing $I^{l}$ immediately before the corresponding text establishes an
explicit association among speaker identity, utterance-level expression, and
linguistic content. Both $I^{g}$ and $I^{l}$ are represented as
natural-language token sequences, enabling open-ended and compositional control
through the existing language-modeling interface. As in JoyVoice, the
autoregressive hidden representations associated with $S$ are further used
to condition the acoustic generation module.

\subsubsection{Paralinguistic Speech Modeling}

Paralinguistic behavior is an important component of natural speech interaction. Vocal events can communicate emotion, attitude, hesitation, physical state, and conversational intent beyond the lexical meaning of the text.

We represent ten types of paralinguistic units using explicit special tokens:

\begin{equation}
\begin{split}
\mathcal{P} = \{&
\texttt{[Laughter]},
\texttt{[Sigh]},
\texttt{[Cough]},
\texttt{[Uhm]},
\texttt{[Confirmation\mbox{-}en]},
\texttt{[Question\mbox{-}en]},\\
&
\texttt{[Surprise\mbox{-}ah]},
\texttt{[Question\mbox{-}ah]},
\texttt{[Question\mbox{-}ei]},
\texttt{[Dissatisfaction\mbox{-}hnn]}
\}.
\end{split}
\end{equation}

These symbols are added to the model vocabulary and treated as special tokens rather than ordinary textual descriptions. They can be inserted directly into the input text at the desired positions, enabling the model to jointly learn their semantic context, temporal placement, and acoustic realization.

For example:

\begin{verbatim}
[Sigh]I understand. Let us try another solution.

I did not expect that at all.[Laughter]That was actually quite funny.
\end{verbatim}

This representation provides an explicit interface for controlling localized non-lexical vocal events while preserving the surrounding linguistic content. Local instructions govern the overall expressive realization of an utterance,
whereas paralinguistic tokens provide position-specific control over localized
non-lexical vocal events. These two interfaces are complementary.

\subsubsection{Low-Cost Target-Voice Adaptation through Style Borrowing}

Collecting expressive target-speaker recordings at scale is expensive. Studio recordings usually provide clean and consistent timbre, but cover only a limited range of emotional states, prosodic patterns, speaking rates, and paralinguistic behaviors. Exhaustively recording all combinations of these attributes is impractical.

We address this problem through a voice-conversion-based SFT data construction strategy. The central idea is to borrow performances rather than voices.

Movies and conversational media provide a large collection of naturally occurring performances, including excitement, hesitation, anger, relief, whispering, shouting, rapid speech, restrained emotion, and various paralinguistic behaviors. We use voice conversion to transfer these expressive performances into the target assistant voice.

Given an expressive source utterance $x^{s}$, its speaking instruction $I$, and a target-speaker reference $r^{t}$, the converted speech can be represented as

\begin{equation}
\widetilde{x}^{t}
=
\mathrm{VC}
\left(
x^{s},
r^{t}
\right),
\end{equation}

where $\widetilde{x}^{t}$ is expected to retain the major expressive characteristics of the source utterance while transferring its speaker identity toward the target voice.

The resulting training sample is formulated as

\begin{equation}
\left(
I^{l},
T,
\widetilde{x}^{t}
\right),
\end{equation}

where $I^{l}$ describes the expressive realization of the converted utterance and $T$ denotes its linguistic content. In this way, diverse performances from different speakers are reinterpreted through a single, consistent assistant voice.

The converted samples constitute the instruction-aware SFT corpus for the
target assistant voice. Quality control is applied to preserve transcription
consistency, target-speaker similarity, expressive relevance, and general audio
quality. In this way, diverse performances produced by different source
speakers are reinterpreted through a consistent assistant identity.

This strategy enables low-cost expressive adaptation without requiring the
target speaker to record every combination of emotion, speaking rate, vocal
effort, phonation, and paralinguistic behavior. The resulting speaker-specific
SFT model serves as the Talker in JoyAI-Talker, preserving a fixed and
recognizable assistant identity while retaining the local instruction-following
capabilities acquired during foundation training.

\subsubsection{Open-Ended Instruction Control}

The current Talker uses open-ended natural-language instructions rather than a closed set of mandatory categorical labels. The instructions may describe one or multiple acoustic attributes with different levels of detail.

The supported instruction dimensions mainly include:

\begin{itemize}
    \item \textbf{Emotion and Affective Expression:}
    Descriptions of primary emotion, emotional intensity, emotional nuance, attitude, and communicative intention, such as happy, sad, angry, excited, surprised, fearful, and so on.

    \item \textbf{Speaking Rate and Rhythm:}
    Descriptions of speaking rate and rhythmic characteristics, such as
extremely fast, fast, normal, slow, extremely slow, and so on.
    
    \item \textbf{Loudness and Vocal Effort:}
    Descriptions of loudness and vocal effort, such as shouting, loud,
normal, soft, whispering, and so on.

    \item \textbf{Timbre and Phonation:}
    Descriptions of temporary timbre characteristics and phonation styles, such as clear, bright, breathy, hoarse, tense, relaxed, and so on.

    \item \textbf{Paralinguistic Behavior:}
    Control of the ten types of paralinguistic units introduced above.

    \item \textbf{Compositional Expression:}
    Natural-language descriptions combining multiple acoustic attributes, such as emotion, speaking rate, loudness, timbre, scene, paralinguistic behavior, and so on.
\end{itemize}

Representing instructions in natural language provides a more flexible control interface than fixed categorical labels. It allows expressive attributes to be described with varied wording and enables multiple control dimensions to be combined within a single instruction.

\subsubsection{Streaming Speech Generation}

The Talker supports streaming text input and streaming speech output. As the Thinker incrementally produces a textual response, the Talker can begin synthesizing the available content without waiting for the complete response.

During streaming inference, the local instruction is provided before its associated text and is preserved as a complete conditioning unit. This is particularly important for natural-language instructions and paralinguistic special tokens, whose meanings may be weakened if they are separated from the corresponding text span.

The streaming interface dynamically organizes incoming text into generation units while maintaining the association among local instructions, paralinguistic events, and response content. This enables low-latency speech generation while preserving the expressive intention of the response.

\subsubsection{Thinker--Talker Interface}

Because the deployed Talker uses a fixed target-speaker voice, no global
speaker instruction is required during inference. The Thinker produces the
textual response together with an optional local speaking instruction derived
from its empathetic understanding of the user, dialogue context, and intended
communicative behavior. The instruction is enclosed by
\texttt{<instruct>} and \texttt{</instruct>} and is placed before the
corresponding response text.

When an explicit instruction is provided, the Talker uses it to control the
utterance-level acoustic realization while preserving the target assistant
identity. When no instruction is provided, the Talker infers an appropriate
speaking style from the linguistic and semantic context. Position-specific
paralinguistic tokens may additionally be inserted into the response text to
control localized vocal events.

This interface separates empathetic reasoning from acoustic realization. The
Thinker determines response content and communicative intent, while the Talker
specializes in speaker-consistent, expressive, and low-latency speech
generation.

\subsection{Duplex: State-Driven Semantic-Gated Full-Duplex Interaction}
\begin{figure}[t]
\centering
\includegraphics[width=1.0\linewidth]{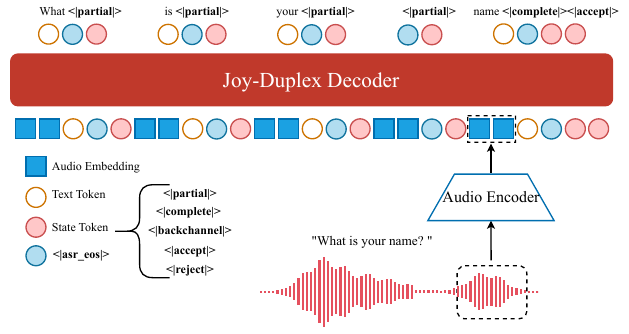}
\caption{\textbf{Unified token-interleaved streaming mechanism and state transitions of the Joy-Duplex Decoder.} The streaming Audio Encoder processes continuous user audio waveforms within a sliding window. The Joy-Duplex Decoder receives continuous Audio Embeddings and autoregressively decodes interleaved Text Tokens, boundary tokens (\texttt{<|asr\_eos|>}), and State Tokens within a single unified sequence. 
}
\label{fig:joy_duplex_framework}
\end{figure}

To enable natural, real-time full-duplex speech interaction without compromising the capabilities of the backbone half-duplex large speech dialogue model (SDM), we depart from the computationally heavy end-to-end parallel dual-stream architectures~\cite{defossez2024moshi, thinkingmachines2026interactionmodels} and propose \textbf{Joy-Duplex}, a decoupled, state-driven modular duplex framework. As illustrated in Fig.~\ref{fig:joy_duplex_framework}, this plug-and-play design equips pre-existing half-duplex systems with robust full-duplex capabilities without modifying their core backbone parameters, facilitating flexible deployment and scaling~\cite{25easyturn,yan2026soulx,26fastturn}.

\subsubsection{Joy-Duplex Architecture}
The Joy-Duplex framework isolates real-time turn-taking control from core response generation, deploying a lightweight external duplex predictor to govern the downstream half-duplex SDM. As illustrated in Fig.~\ref{fig:joy_duplex_framework}, the architecture consists of a streaming Audio Encoder and an interleaved Joy-Duplex Decoder:

\begin{itemize}
    \item \textbf{Streaming Audio Encoder}: 
    To process continuous user speech with low algorithmic latency, we employ a streaming speech encoder that slices the input waveform into non-overlapping 160 ms chunks during inference. To enhance the model's robustness and generalization across various latency requirements, the encoder is trained using a dynamic chunk-based attention strategy~\cite{zhang2020unified}, where the chunk size is dynamically randomized during the training phase. The attention receptive field of each frame is constrained to only look backward at historical context and within the current chunk. The resulting continuous Audio Embeddings are then aligned via a linear projector and directly fed into the decoder.
    
    \item \textbf{Joy-Duplex Decoder}: The core backbone is a 1.7B-parameter decoder-only language model. The Joy-Duplex Decoder accepts the continuous Audio Embeddings alongside historical context, and autoregressively decodes interleaved Text Tokens and State Tokens within a single unified sequence. Notably, to establish explicit structural boundaries between speech transcripts and interaction states, each predicted text segment is strictly terminated with a dedicated \texttt{<|asr\_eos|>} token (represented as blue circles in Fig.~\ref{fig:joy_duplex_framework}). This boundary token explicitly marks the completion of incremental transcription for the current frame, immediately after which the decoder predicts the corresponding interaction State Tokens.
\end{itemize}

\subsubsection{Interleaved Sequence Modeling}
We formulate streaming duplex state control as a token-interleaved sequence modeling task under incrementally available acoustic observations. 
Let $A_t$ denote the continuous Audio Embeddings extracted from the $t$-th chunk, $T_t = \{w_1, w_2, \dots, \texttt{<|asr\_eos|>\}}$ represent the corresponding streaming Text Tokens terminated by the \texttt{<|asr\_eos|>} token, and $S_t$ denote the predicted State Tokens. At each step $t$, conditioned on the historical context $\mathcal{H}_{t-1} = \{A_{<t}, T_{<t}, S_{<t}\}$, the joint probability of predicting the current frame's text and state sequence is factorized as:
\begin{equation}
    P(T_t, S_t \mid A_{\le t}, \mathcal{H}_{t-1}) = P(T_t \mid A_{\le t}, \mathcal{H}_{t-1}) \cdot P(S_t \mid A_{\le t}, T_{\le t}, \mathcal{H}_{t-1})
\end{equation}
As illustrated in Fig.~\ref{fig:joy_duplex_framework}, when the user is speaking, the decoder first predicts the aligned Text Tokens $T_t$ and closes the text segment with the \texttt{<|asr\_eos|>} token (e.g., ``What'' $\rightarrow \texttt{<|asr\_eos|>}$), followed immediately by the State Token $S_t$ (e.g., $\texttt{<|partial|>}$) to capture instantaneous semantic and interaction status. Even during speech pauses or acoustic non-speech chunks where no text tokens are generated, the decoder outputs $\texttt{<|asr\_eos|>} \rightarrow \texttt{<|partial|>}$ to securely hold the turn. Only when a complete syntactic boundary is established (e.g., ``What is your name?''), the sequence generates $\text{name} \rightarrow \texttt{<|asr\_eos|>} \rightarrow \texttt{<|complete|>} \rightarrow \texttt{<|accept|>}$, unblocking the semantic gate to trigger the downstream response.

The downstream SDM remains in a gated standby mode, activated for response generation if and only if the Joy-Duplex Decoder emits an explicit \texttt{<|accept|>} token. To enhance deployment safety in multi-speaker or noisy environments, a lightweight Speaker Verification (SV) plugin can run in parallel alongside the encoder for identity matching. This ensures that the SDM is exclusively triggered by the authorized speaker, forming a secure, modular, and robust full-duplex pipeline.

\subsubsection{Interactive State Vocabulary}
Rather than relying on frame-level acoustic energy which often causes false interruptions during natural hesitations, Joy-Duplex introduces five dedicated state tokens directly integrated into the decoding stream to establish a semantically-informed interaction state machine:
\begin{itemize}
    \item \textbf{\texttt{<|partial|>}}: Continuously generated during ongoing active speech, signaling that the user's semantic intent is still developing. This instructs the downstream SDM to remain in standby mode.
    \item \textbf{\texttt{<|complete|>}}: Triggered immediately when the incremental transcription reaches a coherent syntactic and semantic boundary (e.g., transcribing the end of ``...your name'' $\rightarrow$ \texttt{<|complete|>}). This serves as a precursor for the final interactive gating decision.
    \item \textbf{\texttt{<|backchannel|>}}: Emitted when the user provides short verbal feedback (e.g., ``uh-huh'', ``yeah''). It allows the system to acknowledge the user's presence without initiating a full conversational turn.
    \item \textbf{\texttt{<|accept|>} / \texttt{<|reject|>}}: The semantic rejection gate. Immediately following the \texttt{<|complete|>} token, the decoder evaluates the final interaction intent. If a valid, system-directed query is confirmed, it yields \texttt{<|accept|>}, which unblocks the gate and triggers the downstream SDM to speak. Conversely, if the completed segment consists of background leakage, off-target speech, or ambient noise, the decoder outputs \texttt{<|reject|>}, which explicitly suppresses the SDM to prevent accidental activations.
\end{itemize}

\section{Experimental Results}
\subsection{Text-to-Text (T2T) Evaluation}
To ensure fair and reproducible evaluation, we adopt a standardized evaluation pipeline. Most benchmarks are evaluated with OpenCompass under greedy decoding for deterministic reporting, while RULER is evaluated using its official repository under a 64K context window to remain consistent with its native leaderboard setting. All models evaluated in this section are tested with chat-format inference templates, including conversational special tokens, to better reflect real-world user-interactive scenarios.

To examine whether JoyAI-Talker preserves the text-domain capabilities of its backbone after speech-text joint training and alignment, we evaluate JoyAI-Talker-SFT and JoyAI-Talker-DPO on 14 benchmarks covering knowledge, reasoning, instruction following, long-context modeling, and STEM capabilities. The results are summarized in Table~\ref{tab:instruct_eval}.

We discuss the results from two perspectives: general knowledge, reasoning, and STEM capabilities, as well as instruction following and long-context modeling.

\begin{itemize}
    \item \textbf{General Knowledge and STEM:} JoyAI-Talker demonstrates competitive logical reasoning and academic proficiency. In mathematical reasoning, JoyAI-Talker-DPO reaches $94.62\%$ on MATH and $94.58\%$ on MATH-500. It achieves stable results in complex, college-level reasoning benchmarks, scoring $54.31\%$ on SuperGPQA and $64.51\%$ on GPQA. General knowledge also remains robust, leading on MMLU-Redux with $90.80\%$ and CMMLU with $86.18\%$.
    
    \item \textbf{Instruction Following and Long-Context:} On instruction-following benchmarks, the model exhibits stable compliance. The SFT stage successfully bridges the conversational gap, leading to scores of $83.62\%$ on IFEval and $31.67\%$ on IFBench. The DPO phase further refines these capabilities, reaching $83.70\%$ and $31.83\%$ respectively, while sustaining long-horizon conversational coherence with a RULER (64K) score of $76.88\%$. This progressive improvement demonstrates that preference optimization effectively steers the policy toward stricter constraint adherence without compromising general reasoning.
\end{itemize}

\begin{table}[ht]
\centering
\caption{Text-to-Text (T2T) benchmark results of JoyAI-Talker across different training stages. Best results are marked in bold. Qwen3-Omni is configured with 32B parameters (3B active). JoyAI-Talker is configured with 48B parameters (3B active).}
\label{tab:instruct_eval}
\begin{tabular}{lccc}
\toprule
\textbf{Task} & \textbf{Qwen3-Omni} & \textbf{JoyAI-Talker-SFT} & \textbf{JoyAI-talker-DPO}  \\
\midrule
\rowcolor[gray]{0.92} \textbf{Knowledge} & & & \\
\quad MMLU-Redux & 89.65 & 89.94 & \textbf{90.80} \\
\quad C-Eval & 87.79 & 87.31 & \textbf{87.97} \\
\quad MMLU & 87.90 & 87.72 & \textbf{88.57} \\
\quad CMMLU & 85.87 & 85.14 & \textbf{86.18} \\
\hline
\rowcolor[gray]{0.92} \textbf{Reasoning} & & & \\
\quad MMLU-Pro & 76.08 & 77.58 & \textbf{80.23}\\
\quad SuperGPQA & 49.20 & 50.59 & \textbf{54.31}\\
\quad AGIEval & 42.87 & 50.92 & \textbf{54.22}\\
\hline
\rowcolor[gray]{0.92} \textbf{Instruction Following} & & & \\
\quad IFEval & 82.81 & 83.62 & \textbf{83.70}\\
\quad IFBench & 26.68 & 31.67 & \textbf{31.83}\\
\hline
\rowcolor[gray]{0.92} \textbf{Long-Context} & & & \\
\quad RULER (64K) & 73.99 & 74.68 & \textbf{76.88}\\
\hline
\rowcolor[gray]{0.92} \textbf{STEM} & & & \\
\quad MATH & 92.04 & 92.60 & \textbf{94.62}\\
\quad MATH-500 & 92.20 & 92.56 & \textbf{94.58}\\
\quad GSM8K & \textbf{95.98} & 95.53 & 95.60\\
\quad GPQA & 58.93 & 60.04 & \textbf{64.51} \\
\bottomrule
\vspace{5pt}

\end{tabular}
\end{table}

\subsection{Speech-to-Text (S2T) Evaluation}

To comprehensively evaluate speech-conditioned capabilities, we assess the SFT and DPO models across 23 benchmarks spanning acoustic perception, cross-lingual translation, QA, and emotional quotient (EQ). The comparative results are summarized in Table~\ref{tab:s2t_instruct_eval}.

\begin{table}[ht]
\centering
\caption{Comparison of Speech-to-Text (S2T) Instruct Model performance between Qwen3-Omni and JoyAI-Talker SFT/DPO checkpoints. Best results are marked in bold.}
\label{tab:s2t_instruct_eval}
\begin{tabular}{lccc}
\toprule
\textbf{Task} & \textbf{Qwen3-Omni} & \textbf{JoyAI-Talker-SFT} & \textbf{JoyAI-talker-DPO} \\
\midrule
\rowcolor[gray]{0.92} \textbf{ASR} ~($\downarrow$) & & & \\
\quad Aishell 1 & 1.04 & 0.87 & \textbf{0.86} \\
\quad Aishell 2 & 2.48 & 2.18 & \textbf{2.17} \\
\quad LibriSpeech (clean) & 1.52 & 1.33 & \textbf{1.32} \\
\quad LibriSpeech (other) & 3.22 & \textbf{2.55} & 2.58 \\
\quad Wenetspeech (net) & 4.64 & 4.47 & \textbf{4.46} \\
\quad Wenetspeech (meeting) & 5.97 & \textbf{5.86} & 5.88 \\
\quad Fleurs (zh) & 2.79 & 2.54 & \textbf{2.48} \\
\quad Fleurs (en) & 3.18 & \textbf{3.05} & \textbf{3.05} \\
\quad GigaSpeech & 11.27 & 10.25 & \textbf{9.39} \\
\quad Tedium & 3.49 & \textbf{2.85} & 2.90 \\
\hline
\rowcolor[gray]{0.92} \textbf{Translation} ~($\uparrow$) & & & \\
\quad Covost2 (en2zh) & 44.80 & \textbf{51.29} & 50.37 \\
\quad Covost2 (zh2en) & \textbf{28.50} & 27.89 & 27.95 \\
\quad FLEURS (en2zh) & 41.83 & \textbf{43.99} & 43.84 \\
\quad FLEURS (zh2en) & 26.56 & \textbf{27.66} & 27.32 \\
\hline
\rowcolor[gray]{0.92} \textbf{QA} ~($\uparrow$) & & & \\
\quad AdvBench & 99.62 & \textbf{99.81} & \textbf{99.81} \\
\quad AlpacaEval & \textbf{4.68} & 4.62 & 4.61 \\
\quad CommonEval & \textbf{4.53} & 4.51 & \textbf{4.53} \\
\quad OpenBookQA & 93.19 & 93.84 & \textbf{94.73} \\
\quad MMSU & 81.82 & 82.40 & \textbf{84.65} \\
\quad SD-QA & 73.96 & 73.96 & \textbf{75.77} \\
\hline
\rowcolor[gray]{0.92} \textbf{EQ} ~($\uparrow$) & & & \\
\quad Gender & 97.59 & \textbf{98.25} & 98.20 \\
\quad Age & 65.20 & 76.90 & \textbf{77.30} \\
\quad Emotion & 77.00 & 78.97 & \textbf{79.57} \\
\bottomrule

\end{tabular}
\end{table}

We analyze the model's performance categorized across these critical domains:

\begin{itemize}
    \item \textbf{Acoustic Perception (ASR):} JoyAI-Talker achieves solid results across various test suites representing different scenarios. At the DPO stage, it reaches competitive Word Error Rates (WER) on Aishell-1 ($0.86\%$), Aishell-2 ($2.17\%$), and LibriSpeech-clean ($1.32\%$), while maintaining stable performance on other challenging speech datasets.
    
    \item \textbf{Cross-Lingual Translation:} The model displays competitive speech-to-text translation capabilities, particularly in english-to-chinese translation tasks where it reaches a CoVoST-2 score of $51.29$ and a FLEURS score of $43.99$, ensuring stable bilingual conceptual transfer.
    
    \item \textbf{Speech Reasoning (QA):} Benefiting from our pretraining paradigm designed to alleviate cognitive degradation, the model's textual cognitive capacity successfully transfers to speech-conditioned queries. In speech QA benchmarks, JoyAI-Talker-DPO achieves solid reasoning performance, scoring $94.73\%$ on OpenBookQA, $84.65\%$ on MMSU, and $75.77\%$ on SD-QA, demonstrating its capability in processing complex speech queries.
    
    \item \textbf{Emotional Quotient (EQ) Perception:} Paralinguistic modeling serves as a foundation for our Persona-Adaptive Empathetic Response (PAER) framework. By incorporating empathetic and paralinguistic datasets, JoyAI-Talker displays stable performance in extracting speaker attributes from raw audio, scoring $98.25\%$ in Gender classification, $77.30\%$ in Age classification, and $79.57\%$ in Emotion classification, providing a foundation for emotionally intelligent conversational feedback.
\end{itemize}

\vspace{3pt}
\subsection{Evaluation of Persona-Adaptive Empathetic Response}
\label{sec:empathy_eval}

\subsubsection{Audio Understanding and Chain-of-Thought Reasoning.}

To establish a foundation for personalized empathy, we evaluate the model's capability in perceiving speaker attributes from raw audio. 

\begin{itemize}
    \item \textbf{Acoustic Perception Evaluation:} We assess gender and age recognition on AIR-Bench, and emotion recognition on MER2025. To capture emotional nuances more precisely, we audited the test set labels for Track 1, transitioning from six discrete emotional categories to a richer "primary emotion + mental state description" format.
    
    \item \textbf{Evaluation Settings:} We evaluate the model under two distinct settings: (1) \textit{Direct}, which measures atomic perception capabilities in isolation, and (2) \textit{CoT}, which evaluates the attributes explicitly generated within the model's intermediate Chain-of-Thought reasoning steps during actual dialogue generation.
    
    \item \textbf{Perception Results Analysis:} As shown in Table~\ref{tab:user_state_perception}, JoyAI-Talker displays solid and stable capabilities:
    \begin{itemize}
        \item Under the \textit{Direct} setting, the model demonstrates stable baseline understanding, reaching $98.2\%$ in gender, $77.3\%$ in age, and $79.6\%$ in emotion recognition.
        \item Under the \textit{CoT} setting, the model successfully retains and utilizes its acoustic perception within a continuous conversational reasoning process, achieving a competitive $77.7\%$ accuracy in age prediction.
    \end{itemize}
\end{itemize}

\begin{table}[ht]
\centering
\small
\caption{Gender and age are evaluated on the original AIR-Bench labels, while emotion is evaluated on our refined MER2025 annotations. \textit{Direct} tests the model's atomic perception capability in isolation, whereas \textit{CoT} evaluates the attributes explicitly generated within the model's chain-of-thought reasoning prior to formulating a dialogue response.}
\label{tab:user_state_perception}
\begin{tabular}{l l ccc}
\toprule
Model & Setting & Gender~($\uparrow$) & Age~($\uparrow$) & Emotion~($\uparrow$) \\
\midrule
Qwen3-Omni & Direct & 97.59 & 65.2 & 77.0 \\
\midrule
JoyAI-Talker & Direct & \textbf{98.2} & 77.3 & \textbf{79.6} \\
JoyAI-Talker & CoT & 89.3 & \textbf{77.7} & 72.8 \\
\bottomrule
\end{tabular}
\end{table}

\subsubsection{Empathetic Dialogue Generation.}

We assess the final generated responses to evaluate the model's empathetic and persona-adaptive response quality, with the acoustic realization evaluated separately.

\begin{itemize}
    \item \textbf{Evaluation Setup:} We evaluate the final responses using three English subsets derived from EchoMind (Emotion, Age, and Gender). An LLM-as-a-judge paradigm is employed, rating responses on a 1--10 scale across six weighted dimensions: \textit{Basic Logic} ($10\%$), \textit{Knowledge Consistency} ($10\%$), \textit{Diversity} ($10\%$), \textit{Content Logic} ($20\%$), \textit{Interaction} ($20\%$), and \textit{Empathy} ($30\%$).
    
    \item \textbf{Performance Analysis:} As detailed in Table~\ref{tab:empathy_response}, the results highlight the progressive alignment of our framework:
    \begin{itemize}
        \item JoyAI-Talker achieves competitive results across both the overall weighted scores and the dedicated empathy dimensions.
        \item Favorable gains are observed on the EchoMind-Emotion subset. On EchoMind-Age, while the overall weighted score shows a slight increase, the dedicated empathy score displays a notable improvement (climbing to $7.49$).
        \item These targeted gains demonstrate that our decoupled framework effectively concentrates modeling capacity on deep persona adaptation and emotional resonance by seamlessly integrating audio understanding, explicit persona-driven reasoning, and empathetic response generation.
    \end{itemize}
\end{itemize}

\begin{table}[ht]
\centering
\small
\caption{Empathetic-response evaluation based on LLM-as-a-judge. \textit{Weighted Score} denotes the six-dimensional weighted score, while \textit{Empathy Score} denotes the dedicated empathetic-response dimension.}
\label{tab:empathy_response}
\begin{tabular}{l cc cc}
\toprule
\multirow{2}{*}{Benchmark} & 
\multicolumn{2}{c}{Weighted Score} &
\multicolumn{2}{c}{Empathy Score} \\
\cmidrule(lr){2-3} \cmidrule(lr){4-5}
& Qwen3-Omni & JoyAI-Talker & Qwen3-Omni & JoyAI-Talker \\
\midrule
EchoMind-Emotion~($\uparrow$) & 8.91 & \textbf{9.38} & 7.71 & \textbf{8.28} \\
EchoMind-Age~($\uparrow$) & 8.91 & \textbf{9.08} & 7.03 & \textbf{7.49} \\
EchoMind-Gender~($\uparrow$) & 9.47 & \textbf{9.79} & 9.20 & \textbf{9.73} \\
\bottomrule
\end{tabular}
\end{table}

\subsection{Full Duplex Evaluation}
\subsubsection{Evaluation Benchmark}
To verify the efficacy of our proposed framework, we conduct extensive evaluations using the latest established full-duplex conversational standard. We evaluate Joy-Duplex on \textbf{Full-Duplex-Bench v1.5}~\cite{lin2026fullduplexv1.5}, a benchmark designed to assess an audio assistant's categorical behaviors under four typical conversational overlap scenarios:
\begin{itemize}
    \item \textbf{User Interruption}: The user genuinely attempts to take over the floor (e.g., ``Wait, I want to...''). The system must yield the turn immediately and address the new query ($C_{\text{RESPOND}} \uparrow$).
    \item \textbf{User Backchannel}: The user injects short affirmations (e.g., ``uh-huh'', ``yeah'') to signal engagement rather than taking the turn. The system should ignore these cues and continue speaking ($C_{\text{RESUME}} \uparrow$).
    \item \textbf{Talking to Other}: The user addresses a third party in the environment (e.g., ``Hey Grandma...''). The system must perform semantic addressee detection and hold the floor ($C_{\text{RESUME}} \uparrow$, $C_{\text{RESPOND}} \downarrow$).
    \item \textbf{Background Speech}: Ambient far-field speech or environmental noise occurs. The system must remain robust to this acoustic interference and continue its response ($C_{\text{RESUME}} \uparrow$, $C_{\text{RESPOND}} \downarrow$).
\end{itemize}
Following the benchmark protocol, all system behavioral outputs are classified into four categories: $C_{\text{RESPOND}}$ (yielding and responding), $C_{\text{RESUME}}$ (ignoring and continuing), $C_{\text{UNCERTAIN}}$ (addressee checking), and $C_{\text{UNKNOWN}}$ (unclassified behaviors).

\subsubsection{Results and Discussion}
The quantitative results on Full-Duplex-Bench v1.5 are summarized in Table~\ref{tab:fd_bench_eval_comprehensive}. To ensure a state-of-the-art evaluation, in addition to citing baselines from the original benchmark paper~\cite{lin2026fullduplexv1.5}, we leverage the commercial API to locally evaluate and include performance metrics for the released Gemini 3.1 Live model.

\begin{table}[ht]
\centering
\caption{Behavioral response performance on Full-Duplex-Bench v1.5. For each scenario, the primary target metric is denoted with $\uparrow$ for higher is better and the false trigger or action metric with $\downarrow$ for lower is better. Best results are marked in bold.}
\label{tab:fd_bench_eval_comprehensive}
\begin{tabular}{lccccc}
\toprule
\textbf{Scenario / Metric} & \textbf{Freeze-Omni$^*$} & \textbf{Moshi$^*$} & \textbf{GPT-4o$^*$} & \textbf{Gemini 3.1 Live$^\dagger$} & \textbf{Joy-Duplex} \\
\midrule
\rowcolor[gray]{0.92} \textbf{User Interruption} & & & & & \\
\quad $C_{\text{RESPOND}}$~($\uparrow$) & 0.72 & 0.50 & 0.78 & 0.77 & \textbf{0.88} \\
\quad $C_{\text{RESUME}}$~($\downarrow$)  & 0.12 & 0.26 & 0.10 & 0.20 & \textbf{0.07} \\
\hline
\rowcolor[gray]{0.92} \textbf{User Backchannel} & & & & & \\
\quad $C_{\text{RESPOND}}$~($\downarrow$) & 0.07 & 0.02 & 0.03 & 0.02 & \textbf{0.01} \\
\quad $C_{\text{RESUME}}$~($\uparrow$)  & 0.80 & 0.06 & 0.70 & 0.95 & \textbf{0.96} \\
\hline
\rowcolor[gray]{0.92} \textbf{Talking to Other} & & & & & \\
\quad $C_{\text{RESPOND}}$~($\downarrow$) & 0.58 & 0.20 & 0.91 & 0.27 & \textbf{0.17} \\
\quad $C_{\text{RESUME}}$~($\uparrow$)  & 0.25 & 0.19 & 0.02 & 0.66 & \textbf{0.72} \\
\hline
\rowcolor[gray]{0.92} \textbf{Background Speech} & & & & & \\
\quad $C_{\text{RESPOND}}$~($\downarrow$) & 0.62 & 0.21 & 0.93 & 0.28 & \textbf{0.10} \\
\quad $C_{\text{RESUME}}$~($\uparrow$)  & 0.25 & 0.07 & 0.04 & 0.66 & \textbf{0.85} \\
\bottomrule
\multicolumn{6}{l}{\small $^*$ Results are cited directly from the original Full-Duplex-Bench v1.5 paper~\cite{lin2026fullduplexv1.5}.} \\
\multicolumn{6}{l}{\small $^\dagger$ Results are evaluated by ourselves via the latest commercial Gemini Live API.}
\end{tabular}
\end{table}

As presented in Table~\ref{tab:fd_bench_eval_comprehensive}, Joy-Duplex achieves state-of-the-art performance across all four conversational scenarios, demonstrating superior capabilities in both turn-yielding and turn-holding decisions. Driven by our chunk-masked streaming encoder and explicit state transition gating, the system achieves a strong interruption response rate and suppresses erroneous floor-holding. The remaining unresponded interruptions are primarily attributed to a conservative trade-off introduced by our semantic rejection tokens (\texttt{<|reject|>}), which prioritize preventing false activations under ambiguous acoustic conditions.

When encountering non-floor-taking inputs, this conservative gating policy exhibits notable advantages by maintaining appropriate turn-holding behaviors. When users address third parties, Joy-Duplex reduces false responses to 0.17 while maintaining a 0.72 resume rate. Under background speech interference, false activations are suppressed to 0.10 with an 0.85 resume rate. Furthermore, the system seamlessly maintains conversational flow during user backchannels, achieving a high resume rate of 0.96 with a minimal false trigger rate of 0.01. Overall, these evaluations confirm that Joy-Duplex delivers a well-balanced interaction policy by effectively combining prompt turn-yielding with robust semantic rejection.



\subsection{Speech Tool Calling}

Table~\ref{tab:speech_toolcall_bench} reports the performance of JoyAI-Talker on representative speech function calling benchmarks. Compared with two size-matched baselines a text-only model (Qwen3.6-35B) and a multimodal model (Qwen3-omni). JoyAI-Talker achieves the best performance on Speech-ACEBench (Single) while maintaining competitive performance on Speech-BFCL and Speech-SmartInteract. These results demonstrate that JoyAI-Talker effectively transfers text domain agent capabilities to speech interactions, achieving strong overall performance across diverse speech tool-calling tasks.

We attribute these gains to two complementary components. First, high-quality supervised fine-tuning (SFT) agent training data enables the model to acquire strong tool reasoning and function calling capabilities from the text domain. Second, speech oriented agent training data exposes the model to diverse voice-based tool calling scenarios, enabling it to learn the mapping between speech user intents and corresponding tool invocations. These components equip JoyAI-Talker with robust speech agent capabilities and consistently high tool-calling accuracy across a wide range of real-world voice interaction scenarios.

\begin{table}[ht]
\centering
\caption{Performance comparison on speech function calling benchmarks. Bold denotes the best result among the evaluated models. Qwen3.6 is configured with 35B parameters (3B active). Qwen3-Omni is configured with 32B parameters (3B active). JoyAI-Talker is configured with 48B parameters (3B active).}
\label{tab:speech_toolcall_bench}
\small
\begin{tabular}{lccc}
\toprule
\textbf{Benchmark} &
\textbf{Qwen3.6 (Text only)} &
\textbf{Qwen3-Omni} &
\textbf{JoyAI-Talker} \\
\midrule
Speech-ACEBench (Single)~($\uparrow$)    & 96.63 & 92.79 & \textbf{98.56} \\
Speech-ACEBench (Parallel)~($\uparrow$)  & \textbf{61.36} & 38.63 & 44.32 \\
\midrule
Speech-BFCL (Single)~($\uparrow$)        & \textbf{94.81} & 92.21 & 89.62 \\
Speech-BFCL (Parallel)~($\uparrow$)      & 78.76 & \textbf{84.68} & 78.72 \\
\midrule
Speech-SmartInteract~($\uparrow$)        & \textbf{83.08} & 75.90 & 80.39 \\
\bottomrule
\end{tabular}
\end{table}
\section{Conclusion}

In this work, we introduced JoyAI-Talker, a full-duplex speech dialogue system built upon the decoupled Duplex-Thinker-Talker architecture, which separates conversational state coordination, cognitive planning, and expressive speech generation into distinct modular boundaries. Overall, this work introduces several key technological advancements. First, JoyAI-Talker delivers robust foundation model capabilities that natively support advanced conversational voice agents, enabling complex multi-turn speech reasoning, tool-use, and interactive agentic behaviors directly from raw speech queries. Second, through a unified speech-text joint training pipeline spanning pretraining to preference optimization, the system significantly alleviates the common "cognitive degradation" bottleneck to a certain extent, largely preserving the model's fundamental reasoning, STEM, and logical capabilities in the textual domain while expanding into speech. Third, the Persona-Adaptive Empathetic Response (PAER) framework, structured as a hierarchical cognitive pipeline, integrates non-verbal user-attribute perception with Chain-of-Thought (CoT) reasoning to formulate context-adaptive, emotionally intelligent responses. Fourth, the Talker module utilizes a decoupled, instruction-controllable generation paradigm, translating the Thinker's semantic outputs and speaking instructions into speaker-consistent, expressive speech with precise, token-level controls over utterance-level expressions and localized paralinguistic events. Finally, the state-driven, plug-and-play Joy-Duplex framework  interleaves semantic and interaction state tokens directly into the streaming decoding path, acting as a non-intrusive gating engine for real-time turn control. Comprehensive evaluations demonstrate that JoyAI-Talker achieves highly competitive performance across a diverse range of foundational text-to-text (T2T) and speech-to-text (S2T) benchmarks, comparing favorably with state-of-the-art baselines. Notably, on the rigorous Full-Duplex-Bench v1.5, Joy-Duplex delivers a well-balanced interaction policy, achieving a high responsiveness of $0.88$ under user interruptions while keeping false triggers extremely low under background speech, representing a robust and stable paradigm for fluid, natural speech dialogue.
\section{Limitations}
While JoyAI-Talker achieves state-of-the-art performance in highly empathetic voice interaction and robust state-driven full-duplex control, several inherent limitations remain and motivate our ongoing research.

First, the model's perception is currently focused on speech signals and speaker attributes, lacking a dedicated capability for general audio event reasoning and environmental sound comprehension.  In real-world interaction scenarios, non-speech acoustic cues---such as background noise, sirens, alarms, and music---carry critical contextual information.  Future work will prioritize the development of a unified audio-speech encoder and aligned training corpora to enable holistic environmental audio understanding and robust scene-aware reasoning.

Second, the large-scale Reinforcement Learning (RL) training phase for JoyAI-Talker is still in progress.  Without a dedicated RL alignment stage, the model's performance on highly complex, multi-constraint instructions and voice-driven agent invocation can be relatively constrained.  Beyond conventional optimization of speech-text consistency, we plan to further explore the utility of RL.  Specifically, we aim to design a unified speech-text joint reinforcement learning strategy, investigating how RL can be effectively utilized to boost long-term speech reasoning stability, improve complex instruction-following, and refine the execution of voice-driven agentic workflows.

Finally, Joy-Duplex currently operates as a decoupled, state-driven modular framework.  While this design provides robust plug-and-play flexibility and low training overhead, a fully end-to-end dual-stream architecture can achieve favorable performance in minimizing turn-taking latency and realizing continuous, concurrent input-output synchronization.  In future work, we plan to further explore and compare these two paradigms to optimize the trade-offs between modular state-driven coordination and unified, time-aligned dual-stream sequence modeling along a shared temporal axis.

\section*{Authors (alphabetical order of family name)}

\begin{description}
    \item[Core Contributors:] Yinhao Bai, Jinming Chen, Yafeng Chen, Wei Deng, Boya Dong, Nan Duan, Yu Gu, Weisheng Han, Yankun Huang, Ming Ke, Hao Li, Jingdong Li, Xiangyu Liang, Ning Liu, Yuan Liu, Ji Miao, Jiaqi Wang, Qi Wang, Wenchao Wang, Yuxuan Wang, Zhenfang Wang, Zhangyu Xiao, Chao Xue, Hongfei Xue, Fan Yu, Tianyi Zhang, Yuan Zhang, Yuqi Zhang, Lin Zhu
    \item[Contributors:] Haoran Gu, Yiheng Jiang, Jun Wang, Ziteng Wang
\end{description}



\bibliographystyle{unsrt}
\bibliography{references}

\end{document}